\documentstyle[12pt,epsf,captions]{article}
\def\singlespace 
{\smallskipamount=3.75pt plus1pt minus1pt
\medskipamount=7.5pt plus2pt minus2pt
\bigskipamount=15pa plus4pa minus4pt \normalbaselineskip=12pt plus0pt
minus0pt \normallineskip=1pt \normallineskiplimit=0pt \jot=3.75pt
{\def\smallskip {\vskip\smallskipamount}} {\def\medskip
{\vskip\medskipamount}} {\def\bigskip {\vskip\bigskipamount}}
{\setbox\strutbox=\hbox{\vrule height10.5pt depth4.5pt width 0pt}}
\parskip 7.5pt \normalbaselines} 

\def\middlespace
{\smallskipamount=5.625pt plus1.5pt minus1.5pt \medskipamount=11.25pt
plus3pt minus3pt \bigskipamount=22.5pt plus6pt minus6pt
\normalbaselineskip=22.5pt plus0pt minus0pt \normallineskip=1pt
\normallineskiplimit=0pt \jot=5.625pt {\def\smallskip
{\vskip\smallskipamount}} {\def\medskip {\vskip\medskipamount}}
{\def\bigskip {\vskip\bigskipamount}} {\setbox\strutbox=\hbox{\vrule
height15.75pt depth6.75pt width 0pt}} \parskip 11.25pt
\normalbaselines} 

\def\doublespace 
{\smallskipamount=7.5pt plus2pt minus2pt \medskipamount=15pt plus4pt
minus4pt \bigskipamount=30pt plus8pt minus8pt \normalbaselineskip=30pt
plus0pt minus0pt
\normallineskip=2pt \normallineskiplimit=0pt \jot=7.5pt
{\def\smallskip {\vskip\smallskipamount}} {\def\medskip
{\vskip\medskipamount}} {\def\bigskip {\vskip\bigskipamount}}
{\setbox\strutbox=\hbox{\vrule height21.0pt depth9.0pt width 0pt}}
\parskip 15.0pt \normalbaselines}

\newcommand{\be}{\begin{equation}}
\newcommand{\ee}{\end{equation}}
\newcommand{\bea}{\begin{eqnarray}}
\newcommand{\eea}{\end{eqnarray}}
\begin{document}
\begin{center}
\large {\bf Recent direct measurement of the Top quark mass and 
quasi-infrared fixed point} \\ 
\vskip 1in
Biswajoy Brahmachari \\
\end{center}
\begin{center}
Department of Physics and Astronomy\\
University of Maryland, College Park\\
MD-20740, USA.\\
\end{center}
\vskip 1in
{
\begin{center}
 \underbar{Abstract} \\
\end{center}
{\bf {\small
We note that the recent direct measurement of the top quark mass
at $173.3 \pm 5.6 ~{\rm(stat)} \pm 6.2 ~{\rm(syst)}$ by ${\rm D\O}$ 
collaboration severely constrains the theoretically attractive 
infra-red fixed point scenario of the top quark Yukawa coupling in 
supersymmetric GUTs. For one-step unified models the above 
mentioned measurement bounds the arbitrary but experimentally determinable 
parameter $\tan \beta$ to the range $1.3 \le \tan \beta \le 2.1$. Further 
crunch on the top quark mass may determine $\tan \beta$ even more 
accurately within the fixed point scenario. On the other hand an 
experimental value of $\tan \beta > 2.1$ will rule out the fixed point 
scenario bounding $h^2_t(M_X)/4 \pi$ to 0.022 from above.
}}
\newpage
\doublespace
There is a lot of interest among physicists in the possible measurement of 
the top quark mass $m_t$ by the ${\rm D\O}$ group in the vicinity of 173 GeV 
\cite{topmass,topold}. What does this kind of a value mean for various 
theoretical models trying to generate $m_t$ ? In a theoretically 
well-motivated scenario \cite{fixpt} the top quark Yukawa 
coupling, and hence $m_t$, may get fixed at an infrared stable fixed point 
by the structure of the renormalization group equations (RGE) which are a 
coupled set of differential equations for the Yukawa couplings 
$h_i,~{\rm where}~~i=t,b,\tau$ with the gauge couplings. Using RGE 
if we evolve of the top quark Yukawa coupling $h_t$ from a 
large mass scale ($M_X \simeq 10^{16} GeV $) to the scale $m_t$ we obtain a 
universal value of $h_t(m_t)$ for a large domain of values of $h_t(M_X)$. 
The result is noticeable as it shows how the details of the 
possibly complicated symmetry breaking mechanisms at $M_X$ might be 
obliterated by the renormalization group equations, whose fixed point 
structure emerges dominant at low energies. The insensitivity to the 
ultraviolet behaviour is a hallmark of infrared stable fixed points in 
all branches of physics. In this communication we want to test this 
attractive scenario in the light of recent direct measurement of the top 
quark mass by the ${\rm D\O}$ collaboration, who have reported,
\begin{equation}
m_t=173.3 \pm 5.6 ~{\rm(stat)} \pm 6.2 ~{\rm (syst)} ~{\rm GeV}.
\end{equation}
The prediction of the top quark mass in supersymmetric models depends
on an arbitrary parameter $\tan \beta \equiv {\langle H_2 \rangle \over  
\langle H_1 \rangle}$. Where $H_2$ and $H_1$ are the two Higgs doublets
coupling to the top-quark sector and the bottom quark sectors, in 
contrast with the non-supersymmetric case where just one-Higgs doublet 
suffice, prompted by the requirement that the supersymmetric Yukawa 
superpotential has to be an analytic functional \cite{west} of the 
superfields. In one-step supergravity GUT models, the couplings $h_t$, 
$h_b$ and $h_\tau$ are free parameters at the GUT scale, which are to be 
fixed by the low energy Fermion masses. Considering solely the third 
generation 
Yukawa couplings, at low energy the top quark mass satisfys the following 
relation \cite{mtpole} with the top quark Yukawa coupling, $\tan \beta$ 
and the QCD coupling $\alpha_s$; 
\begin{equation}
m^{pole}_t=h_t~(m_t)~174~{\tan \beta \over \sqrt{1+ \tan^2\beta}}
~(1+{ 4 \over 3 \pi}~\alpha_s+ 11.4~{\alpha^2_s\over \pi^2})~
{\rm GeV}. 
\label{mtop} 
\end{equation} 
All entries of this relation except $h_t(m_t)$ are experimentally 
determinate, in principle, eliciting a value of $h_t(m_t)$,  
related via RGE to $h_t(M_X)$. On the other hand if $h_t(M_X)$ is fixed 
by theoretical motivations, Eqn. (\ref{mtop}) can be used \cite{tanb} 
to determine $\tan \beta$, the yet un-known `typical' parameter of 
supersymmetry. The purpose of this brief communication is to 
evaluate the bound on $\tan \beta$ vis-a-vis the latest measurement of 
the top quark mass and to identify it's consequences for the fixed point 
scenario.

We use the two-loop RGE \cite{jones} to evolve the Yukawa as well as the 
gauge couplings. This determines for us the value \cite{as} of 
$\alpha_s=0.125$. Now the bounds on the top quark mass gives the upper 
bounds on $\tan \beta$ which have been plotted in Figure (1). The lower 
bound on $\tan \beta$ comes from the perturbative unitarity arguments 
and it depends on the top quark mass which are the shaded regions in the 
left of Figure (1). The lower bound $\tan \beta > 1.3$ is valid for the 
lower bound on the top quark mass $m_t>161 ~{\rm GeV}$. For the fixed 
point scenario $h^2_t(M_X)/4\pi \simeq 1$, we get the upper bound 
$\tan \beta <2.1$.
\begin{figure}[htb]
\begin{center}
\epsfxsize=11cm
\epsfysize=11cm
\mbox{\hskip 0in}\epsfbox{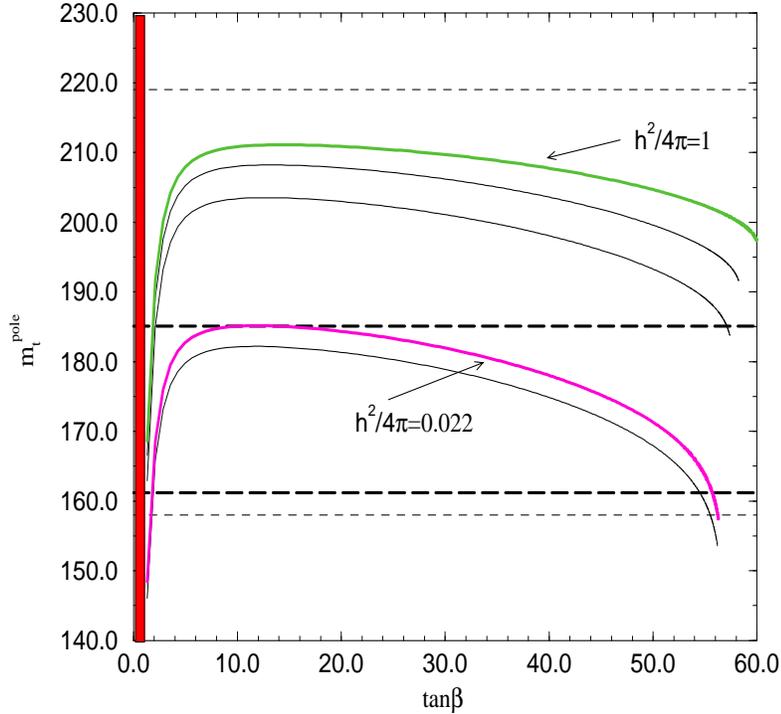}
\caption{The fixed point scenario of the top quark Yukawa coupling
$h^2_t(M_X)/4\pi=1$ is the upper solid line. Lower curves represent lower 
values of the top quark Yukawa coupling at the GUT scale. The band at the 
left is the excluded region in the $\tan \beta$ plane from the 
perturbative unitarity of the top quark Yukawa coupling. We notice
that the horizontal line at 219 GeV (old experimental upper bound) 
does not intersect with the predicted contours.
\label{}} 
\end{center} 
\end{figure} 
We see that apart from being attractive from a theoretical point of view, 
the infra-red fixed point scenario makes definitive predictions of the
measurable parameter $\tan \beta$, and hence, can well be falsified.  
The parameter $\tan \beta$ can be determined from various studies of the
supersymmetric Higgs sector for example \cite{feng} from the heavy 
Higgs scalar production at future $e^+e^-$ colliders or from inclusive
semileptonic B-decays \cite{coar} which depends on the sign of the 
supersymmetric mixing parameter $\mu$ or from the decay of heavy 
charged Higgs ($m_{H^+} >> m_t$) \cite{czar}. In any case if the value 
of $\tan \beta$ is determined to be greater than 2.1, the fixed point 
scenario in the minimal one-step unified models will be excluded.  

For completeness, we note that the fixed point scenario the prediction of 
the b quark mass \cite{lang} is in the range 
$4.29 ~({\rm GeV})< m_b(m_b) < 4.61~({\rm GeV})$, whereas, in the 
case of $h^2_t/4\pi=0.022$ the prediction turns out to
$5.44 ~({\rm GeV})< m_b(m_b) < 6.24~({\rm GeV})$ for all possible variation
of the parameter $\tan \beta$, where the higher end of the b-quark mass 
range corresponds to a lower end of $\tan \beta$ domain.

Now we wish to make a remark on this deceptively simple analysis. What is 
new? The upper-bound on $\tan \beta$ comes from the intersection point 
of the experimental value of the top-quark mass (dashed lines) with the 
predictions. We notice that the previous \cite{topold} experimentally 
allowed range (dotted lines) of the top quark mass was from 158 GeV all 
the way to 219 GeV [${\rm CDF}~~176\pm 18$ GeV and ${\rm D\O}~~199 \pm 20$ 
GeV] the upper experimental bounds did not intersect the theoretical 
predictions. Where as, the recent direct measurements intersect the 
contours of theoretical predictions giving upper bounds on $\tan \beta$ 
corresponding to the point of intersection, which may be worth stressing, 
as we await more accurate determination of the top quark mass.

In conclusion, we have re-evaluated the upper bounds on $\tan \beta$ in 
the context of the recent direct measurement of the top-quark mass by the 
${\rm D\O}$ collaboration. We have shown that if the infra red fixed point 
scenario is correct, $\tan \beta$ should be less than 2.1. On the 
contrary, if $\tan \beta > 2.1$ the fixed point scenario will be ruled 
out bounding the coupling ${h^2_t(M_X)\over 4 \pi} < 0.022$; another 
motivation for the experimentalists to fix the value of $\tan \beta$ in 
supersymmetric theories.

\noindent This work has been supported by a grant from National Science 
Foundation. I would like to acknowledge discussions with Prof. R. N. 
Mohapatra.


\begin{thebibliography}{References}

\bibitem{topmass} ${\rm D\O}$ collaborations, S. Abachi {\it et~al}, 
${\rm D\O}$ collaborations, hep-ex/9703008.

\bibitem{topold} CDF collaborations, F. Abe {\it et~al}, Phys. Rev. Lett. 
{\bf 74}, 2626 (1995); ${\rm D\O}$ collaborations, S. Abachi {\it et~al}, 
Phys. Rev. Lett. {\bf 74}, 2632 (1995).  

\bibitem{fixpt} B. Pendleton and G. G. Ross, Phys. Lett. {\bf B98}, 291, 
(1981); C. T. Hill, Phys. Rev. {\bf D24}, 691, (1981); see also,
M. Carena, C. E. M. Wagner, Nucl. Phys. {\bf B452}, 45 (1995). 

\bibitem{west} For a review see, P. West, ``{\it Introduction to 
supersymmetry and supergravity}'' (World Scientific, Singapore, 1990) 

\bibitem{mtpole} R. Tarrach, Nucl. Phys. {\bf B183}, 384 (1981); 
H. Arason, D. J. Castano, B. Keszthelyi, S. Mikaelian, E.J. Piard, P. 
Ramond, B.D. Wright, Phys. Rev. {\bf D46}, 3945 (1992).

\bibitem{tanb} See reference \cite{berg} and also M. Carena, S. 
Dimopoulos, C. E. M. Wagner, S. Raby, Phys. Rev. {\bf D52}, 4133 (1995).

\bibitem{jones} By D. R. T. Jones Phys.Rev. {\bf D25}, 581, (1982); M. B. 
Einhorn, D. R. T. Jones, Nucl. Phys. {\bf B196}, 475 (1982). Expressions 
of anomalous dimensions are given in, M. Bastero-Gil and B. Brahmachari 
Nucl. Phys. {\bf B482}, 39 (1996).

\bibitem{as} For detailed evaluation of $\alpha_s$ in supersymmetric GUTs 
see for example, J. Bagger, K. Matchev and D. Pierce, Phys. Lett. {\bf B 
348}, 443 (1995); P. H. Chankowski, Z. Pluciennik and S. Pokorski, Nucl. 
Phys. {\bf B 439}, 23 (1995); M. Bastero-Gil and J. P\'erez-Mercader, Nucl 
Phys. {\bf B450}, 21 (1995).                            

\bibitem{berg} See for example: V. Barger, M. S. Berger and P. Ohmann, Phys. 
Rev. {\bf D 47}, 1093, (1993). For a brief summary of the RGE results in 
SUSY GUTs see, V. Barger, M. S. Berger, P. Ohmann and R. N. J. Phillips, 
University of Wisconsin Madison Report No. MAD/PH/803 (unpublished) and 
references therein; Phys. Lett. {\bf B314}, 351, (1993). 

\bibitem{feng} Jonathan. L. Feng, Takeo Moroi, hep-ph/9612333. 

\bibitem{coar} J.A. Coarasa, R. A. Jimenez, J. Sola, hep-ph/9701392. 

\bibitem{czar} A. Czarnecki, J. L. Pinfold, Phys. Lett. {\bf B328}, 427 
(1994).
         
\bibitem{lang} 
J.E. Bjorkman, D.R.T. Jones; Nucl. Phys. {\bf B259}, 533 (1985); B. 
Ananthnarayan, G. Lazarides and Q. Shafi, Phys. Rev. 
{\bf D 44}, 1613 (1990); P. Langacker, N. Polonsky, Phys. Rev. {\bf D49}, 
1454 (1994). R. Rattazzi, U. Sarid, L. J. Hall, SU-ITP-94-15, Presented 
at 2nd IFT Workshop on Yukawa Couplings and the Origins of Mass, Feb. 
11-13, 1994, Gainesville, Florida. Also, Arason {\it et~al}, in Ref. 
\cite{mtpole}. Nir Polonsky, Phys. Rev. {\bf D54}, 4537 (1996). 
\end{thebibliography}
\end{document}